\documentclass{article}
\usepackage[margin=1in]{geometry}
\usepackage[onehalfspacing]{setspace}
\usepackage{amsmath,mathtools}
\usepackage{pgfplots}
\usepackage{subcaption}
\usepackage{booktabs}
\usepackage[braket, qm]{qcircuit}
\usepackage{graphicx}
\usepackage{authblk}
\usepackage{xcolor}
\usepackage[linesnumbered,ruled,vlined]{algorithm2e}
\usepackage{qcircuit}

\SetCommentSty{mycommfont}

\SetKwInput{KwInput}{Input}                
\SetKwInput{KwOutput}{Output}              
\SetKwProg{Fn}{Function}{}{}

\usepackage[english]{babel}

\usepackage{pdflscape}
\usepackage{fancyvrb}
\usepackage{calc}
\usepackage{rotating}
\usepackage{pgf,tikz}
\usepackage{mathrsfs}
\usetikzlibrary{arrows}
\usepackage{mathtools}

\usepackage{listings}
\usepackage[]{qcircuit}
\usepackage{braket}
\usepackage{mathtools}
\usepackage{varwidth}
\usepackage{caption}
\usepackage{subcaption}
\usepackage{graphicx}
\usepackage[export]{adjustbox}

\usepackage{float}

\makeatletter
\def\paragraph{\@startsection{paragraph}{4}%
	\z@\z@{-\fontdimen2\font}%
	{\normalfont\bfseries}}
\makeatother

\usepackage{graphicx}
\usepackage{pgffor}
\usepackage{tikz,tikz-cd}
\usetikzlibrary{backgrounds,fit,decorations.pathreplacing}  
\usepackage{booktabs}
\usepackage{enumerate}

\usepackage{amssymb, amsmath, amsthm}

\usepackage{xypic}
\usepackage{scalerel}
\usepackage{ulem}

\usepackage{graphicx}              
\usepackage{amsmath}               
\usepackage{amsfonts}              
\usepackage{amsthm}                
\usepackage{xkeyval}               
\usepackage{amssymb}
\usepackage{enumerate}             
\usepackage{color}
\usepackage{breqn}                 
\usepackage{bm}                    
\usepackage{cite}

\allowdisplaybreaks[1]
\usepackage{nicefrac}
\usepackage{booktabs}

\usepackage{kpfonts}
\usepackage{stackengine}
\usepackage{calc}
\newlength\shlength
\newcommand\xshlongvec[2][0]{\setlength\shlength{#1pt}%
	\stackengine{-5.6pt}{$#2$}{\smash{$\kern\shlength%
			\stackengine{7.55pt}{$\mathchar"017E$}%
			{\rule{\widthof{$#2$}}{.57pt}\kern.4pt}{O}{r}{F}{F}{L}\kern-\shlength$}}%
	{O}{c}{F}{T}{S}}

\usepackage{hyperref}
\hypersetup{
	colorlinks   = true,    
	citecolor    = red      
}

\newcommand{\RN}[1]{%
	\textup{\uppercase\expandafter{\romannumeral#1}}%
}

\allowdisplaybreaks

\newcommand{\meqref}[1]{\text{Eq}.~\eqref{#1}}
\newcommand{\mref}[1]{Sec.~$ \!\ref{#1} $}
\newcommand{\mfig}[1]{Fig.~$ \!\ref{#1} $}

\def\<{\langle}
\def\>{\rangle}

\numberwithin{equation}{section}

\begin{document}

	\title{A generalization of Bernstein–Vazirani algorithm with multiple secret keys and a probabilistic oracle}

	\author[1]{Alok Shukla \thanks{Corresponding author.}}
	\author[2]{Prakash Vedula}
	\affil[1]{School of Arts and Sciences, Ahmedabad University, India}
	\affil[1]{alok.shukla@ahduni.edu.in}
	\affil[2]{School of Aerospace and Mechanical Engineering, University of Oklahoma, USA}
	\affil[2]{pvedula@ou.edu}
	

	\maketitle

	\begin{abstract}

		A probabilistic version of the Bernstein-Vazirani problem (which is a generalization of the original Bernstein-Vazirani problem) and a quantum algorithm to solve it are proposed. The problem involves finding one or more secret keys from a set of multiple secret keys (encoded in binary form) using a quantum oracle. From a set of multiple unknown keys, the proposed quantum algorithm is capable of (a) obtaining any key (with certainty) using a single query to the probabilistic oracle and (b) finding all keys with a high probability (approaching 1 in the limiting case). In contrast, a classical algorithm will be unable to find even a single bit of a secret key with certainty (in the general case). 
		Owing to the probabilistic nature of the oracle, a classical algorithm can only be useful in obtaining limiting probability distributions of $ 0 $ and $ 1 $ for each bit-position of secret keys (based on multiple oracle calls) and this information can further be used to infer some estimates on the distribution of secret keys based on combinatorial considerations. 
		For comparison, it is worth noting that a classical algorithm can be used to exactly solve the original Bernstein-Vazirani problem (involving a deterministic oracle and a single hidden key containing $n$ bits) with a query complexity of $\mathcal{O}(n)$.
		An interesting class of problems similar to the probabilistic version of the Bernstein-Vazirani problem can be construed, where quantum algorithms can provide efficient solutions with certainty or with a high degree of confidence and classical algorithms would fail to do so. 	
	\end{abstract}

	\section{Introduction}\label{sec:intro}

In recent years, many interesting quantum and hybrid classical-quantum algorithms  \cite{nielsen2002quantum,deutsch1992rapid,bernstein1993quantum,grover1997quantum,simon1997power,shor1999polynomial,harrow2009quantum,wittek2014quantum,childs2020quantum,lloyd2020quantum,cerezo2021variational,shukla2019trajectory,SHUKLA2023127708,Yan2016,Shukla2022} have been proposed with applications in a wide range of fields  that often offer computational advantages over their classical counterparts. For all of these algorithms corresponding classical solutions can be obtained with certainty. In this work we will consider a problem (based on a generalization of the Bernstein–Vazirani problem  \cite{bernstein1993quantum}) and a quantum algorithm to solve it with certainty, while any classical algorithm will fail to obtain a solution with certainty.

Bernstein–Vazirani algorithm  \cite{bernstein1993quantum} is often presented as a textbook example to demonstrate the superiority of a  quantum algorithm over its classical counterpart.  The Bernstein-Vazirani algorithm belongs to the BQP (bounded-error quantum polynomial time) complexity class, which represents a class of decision problems that can be solved in polynomial time with a bounded probability of error. We recall that Bernstein–Vazirani algorithm  finds the secret string $ a $ with $ a \in \{0,1\}^{n} $, given a function $ {\displaystyle f\colon \{0,1\}^{n}\rightarrow \{0,1\}} $  defined as $  f(a)= s \cdot a $. Here we are using the convention that $ x \cdot y $ denotes the bit-wise dot product of  $ x $ and $ y $ modulo $ 2 $, i.e.,  $x \cdot y =  x_{0}y_{0} + x_{1}y_{1}+ \ldots + x_{n-1}y_{n-1} \mod 2 $ where $ x = \sum_{k=0}^{n-1} \, x_k 2^k  $  and $ y = \sum_{k=0}^{n-1} \, y_k 2^k  $. Access to a black-box oracle implementing the function $ f $ is assumed. Classically, at least $ n $ queries to the oracle are needed to determine the secret key, whereas Bernstein–Vazirani algorithm needs only one call to the oracle. 
	
\begin{figure}[H]
		\centering
		\includegraphics[scale=1.5]{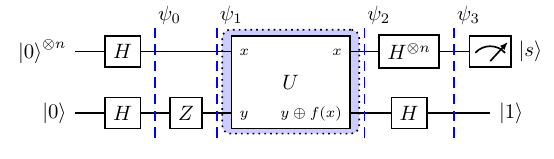}
		\caption{Quantum circuit for Bernstein–Vazirani algorithm. } \label{fig:BV}
	\end{figure}

We also note that the original Bernstein–Vazirani algorithm can be implemented using the quantum circuit shown in \mfig{fig:BV}, where the action of the unitary gate $ U $  is given by
	\begin{equation}\label{eq_action_unitary}
		U \ket{x} \otimes \ket{y} = \ket{x} \otimes \ket{y \oplus f(x)},
	\end{equation}
	where $ f(x) = s \cdot x$ and the secret key $ s\in \{0,1\}^{n}  $.

In this article, we present a generalization of Bernstein–Vazirani algorithm in the presence of multiple secret keys.
In particular, we will describe a probabilistic version of Bernstein–Vazirani problem along with a quantum algorithm to solve it (see \mref{sec:pbv_algo}). 

Many well-known problems could be solved by both classical and quantum algorithms, with quantum algorithms offering computational advantages over their classical counterparts (exponential advantages in some cases). 
In contrast, it is interesting to note that, the probabilistic version of the Bernstein-Vazirani problem belongs to a class of problems,  where quantum algorithms can be used to obtain efficient solutions with certainty (with high probability, approaching $ 1 $ in the limiting case), while classical algorithms fail to obtain accurate solutions.

	\section{Probabilistic Bernstein–Vazirani Algorithm}
	\label{sec:pbv_algo}
	First we describe our probabilistic version of Bernstein–Vazirani problem. For the ease of reference, henceforth we will refer to this problem as the probabilistic Bernstein–Vazirani problem.

	\subsection{Problem Statement }	
	In the probabilistic Bernstein–Vazirani problem, the goal is to find multiple secret keys  (or strings), whereas the classical version involves finding only one secret key.  Also, unlike the use of a deterministic oracle in the classical version, a  probabilistic black-box oracle $ O_{k} $ will be considered here. Assume that the functions $ {\displaystyle f_i \colon \{0,1\}^{n}\rightarrow \{0,1\}} $ are defined as $ f_i(x) = s_i \cdot x $,  with distinct secret keys $ s_i \in \{0,1\}^{n}  $, for  $ i=0,\,1, \, \ldots \, k-1 $. Here $ 1 \leq k \leq 2^n $. The oracle $ O_k $ is such that for an input $ x $, it returns any one of the outputs $ f_0 (x), \, f_1 (x), \, \ldots \,, f_{k-1}(x)$ with equal probability of $ \frac{1}{k} $.  
	Our probabilistic version of Bernstein–Vazirani problem involves two sub-problems as described below.
	\begin{enumerate}[(a)]
		\item Find any one secret key with certainty.
		\item Find all the secret keys $ s_0 $, $ s_1 $, $\ldots  $, $ s_{k-1} $ with high probability. 
	\end{enumerate} 
	Note that for both the above sub-problems, it is desired that as few calls to the oracle $ O_{k} $ are made as possible. 
	For the sub-problem (b), by the phrase ``high probability'', it is meant that in the limiting case the probability of finding all the secret keys should approach $ 1 $, as the number of oracle queries approach infinity. Further, we do not require the secret keys to be distinct, i.e., two or more secret keys can be identical. In case all the secret keys are identical the above problem reduces to the original Bernstein–Vazirani problem.

	\subsection{Algorithm}	\label{sec:algo}
\begin{figure}[H]
		\centering
		\includegraphics[scale=0.93]{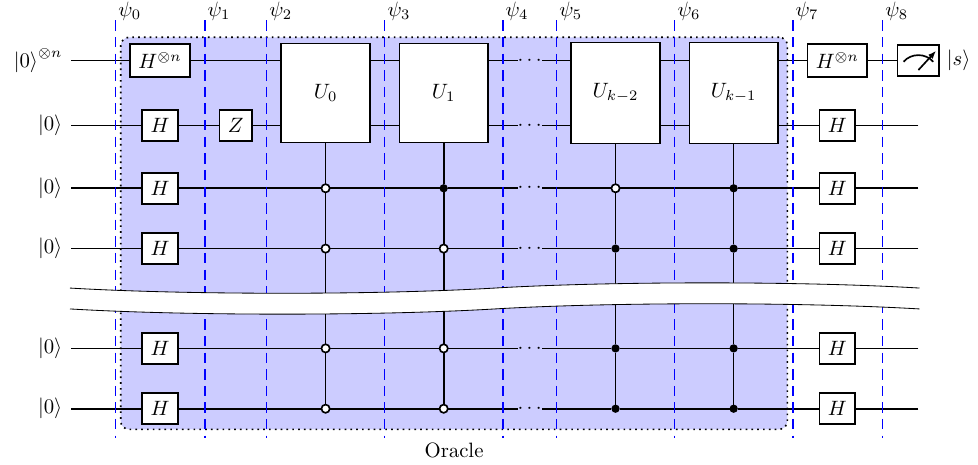}
		\caption{Circuit for implementation of the probabilistic version of the Bernstein-Vazirani algorithm for deciphering $k = 2^r$ secret keys (each containing $n$ bits). The shaded region represents the oracle.} \label{fig:main}
	\end{figure}

	The quantum circuit given in \mfig{fig:main} that uses the above mentioned oracle can be employed to solve the probabilistic Bernstein–Vazirani problem described earlier.  
	The shaded region in \mfig{fig:main} represents  a quantum oracle $ O_{k} $. The implementation details of this oracle will be described later in  \mref{Sec:Oracle}. 		The action of the oracle $ O_{k} $ on $ \ket{0}^{\otimes n} $ is given by 

	\begin{equation}\label{eq_oracle}
		O_{k} \ket{0}^{\otimes n}  =    \frac{1}{\sqrt{k}}  \,  \sum_{i=0}^{k-1} \, \ket{\Psi_i},
	\end{equation}
where  \begin{equation}\label{eq_phi_i}
	\displaystyle {\ket{\Psi_i} = \frac{1}{\sqrt{N}} \, \sum_{x =0}^{N-1} \, (-1)^{f_i(x)} \ket{x}},
\end{equation}
for $ i = 0, \, 1,\, \ldots \,, k-1$, and where $ N =2^n $. For ease of illustration, here we assume that $ k = 2^j $ for some non-negative integer $ j $. However this restriction on $ k $ is not required as noted in  \mref{Sec:Oracle}.
The quantum state given in \meqref{eq_oracle} corresponds to the quantum state of the top $ n $-qubits of  $ \ket{\psi_7} $ as shown in \mfig{fig:main}.

	The output state of the top $ n $ qubits, in the circuit shown in \mfig{fig:main}, is given by 
	\begin{align}
		&H^{\otimes n}  O_{k} \ket{0}^{\otimes n}  
		 = 	H^{\otimes n}  \left[  \frac{1}{\sqrt{k}}  \, \left(  \ket{\Psi_0} + \ket{\Psi_1} + \ldots +  \ket{\Psi_{k-1}}  \right)  \right]  \nonumber \\  
		& 	= H^{\otimes n}  \left[  \frac{1}{\sqrt{k}}  \, \left(  \frac{1}{\sqrt{N}} \, \sum_{x =0}^{N-1} \, (-1)^{s_0 \cdot x} \ket{x}   +  \frac{1}{\sqrt{N}} \, \sum_{x =0}^{N-1} \, (-1)^{s_1 \cdot x} \ket{x}    + \ldots +  \frac{1}{\sqrt{N}} \, \sum_{x =0}^{N-1} \, (-1)^{s_{k-1} \cdot x} \ket{x} \right)  \right]  \nonumber \\
		& 	=     \frac{1}{N \sqrt{k}}  \, \left(   \sum_{y =0}^{N-1}  \sum_{x =0}^{N-1} \, (-1)^{s_0 \cdot x + y \cdot x} \ket{y}   +    \sum_{y =0}^{N-1} \sum_{x =0}^{N-1} \, (-1)^{s_1 \cdot x + y \cdot x} \ket{y}  + \ldots +   \sum_{y =0}^{N-1} \sum_{x =0}^{N-1} \, (-1)^{s_{k-1} \cdot x + y \cdot x} \ket{y}\right)   \label{Eq_super_sum}
	\end{align}
	If $ y =s_i $, then $ y \oplus s_i = 0 $, which means 
	$ \frac{1}{ N }   \,   \sum_{x =0}^{N-1} \, \left( (-1)^{(s_i \oplus y) \cdot x }  \right) \ket{y}  = \frac{1}{ N }      \sum_{x =0}^{N-1} \, \left( (-1)^{(0) \cdot x }  \right) \ket{s_i}  =  \ket{s_i}$. Hence, if  $ y \neq s_i $, then the corresponding coefficient of $ \ket{y}$ must be $ 0 $. It follows that  $  \frac{1}{N} \, \sum_{y =0}^{N-1}  \sum_{x =0}^{N-1} \, (-1)^{s_i \cdot x + y \cdot x} \ket{y} = \ket{s_i} $,  for $ i=0,\,1, \, \ldots \, k-1 $. We note that a similar computation is also present in the original Bernstein–Vazirani algorithm.  
	Further, from  \meqref{Eq_super_sum}, we obtain 
	\begin{equation}\label{eq_last_result}
		H^{\otimes n}  O_{k} \ket{0}^{\otimes n} = \frac{1}{\sqrt{k}} \left( \ket{s_0} + \ket{s_1} + \, \ldots \,  + \ket{s_{k-1}}  \right).
	\end{equation}

	It follows that a measurement (see \mfig{fig:main}) returns any one of the secret keys $ s_0 $ or $ s_1 $, $ \ldots $, or $ s_{k-1} $ with equal probability $ \frac{1}{k} $, in the classical query register. Hence only one oracle query is needed to find any one of the $ k $ secret keys. Note that this solves the sub-problem $ (a) $ of the probabilistic Bernstein–Vazirani problem with just one oracle query. We further note that repeated application of this quantum algorithm will eventually allow us to determine all the secret keys. In the following, we will provide detailed comparisons of quantum and classical approaches for solutions of both the sub-problems (i.e., sub-problems $ (a) $ and $ (b) $) of the probabilistic version of the Bernstein–Vazirani problem.
	
	\subsubsection{Sub-problem $ (a) $: Find any one secret key}
	
	We note here that classically it is impossible to deterministically solve the sub-problem $ (a) $ of the probabilistic Bernstein–Vazirani problem for all cases involving at least two distinct secret keys. In order to see this, first we observe that in the case of  the original Bernstein–Vazirani problem, the most efficient classical method for obtaining the $ n $-bit secret key $ s $ (with decimal representation $ s = \sum_{q=0}^{n-1}  s(q) 2^q$), is by evaluating the function $ f(x) $, $ n $-times with inputs $ x = 2^q $, for all $ q \in \{0, \,1,\, \ldots,\,n-1 \} $. This returns the bit $ s(q) $ of the secret key $ s $ for each query. Recall that the original Bernstein–Vazirani (quantum) algorithm requires only one query to the oracle, while the classical solution requires at least with $ n $ oracle queries. Next we consider the probabilistic Bernstein–Vazirani problem and assume that the secret keys $ s_i $ and $ s_j $ are distinct. It means  $ s_i(q) \neq s_j(q) $ for some $ q \in \{0, \,1,\, \ldots,\,n-1 \} $. Extension of the classical approach (discussed above) for solution of sub-problem $ (a) $ of the probabilistic Bernstein–Vazirani problem, will fail to determine even a single bit of the secret key with certainty. The reason for this claim is that a single query to the oracle ($ O_{k} $) of the probabilistic Bernstein–Vazirani problem with $ x = 2^q $ could result in a bit $ \tilde{s}(q) $, where $ \tilde{s}$ is probabilistically chosen from  $\{s_0,\,s_1,\,\ldots,\,s_{k-1}\}$.
	It means $ \tilde{s}(q) $ originates from any of the $ k $ secret keys $ s_i $ (with the decimal representation  $ s_i =  \sum_{q=0}^{n-1}  s_i(q) 2^q$). The probabilistic nature of the oracle makes it impossible to associate the output bit $ \tilde{s}(q) $ to the secret key $ s_i $ from which it originates  (i.e., $ i $ can not be determined, even when the bit $ \tilde{s}(q) $ is known by suitably querying the oracle with the input $ x = 2^q $). Therefore, it is not possible to determine even one secret key with certainty using any classical algorithm. This provides us with an interesting case where it is impossible to obtain a deterministic solution (to the sub-problem $ (a) $) using a classical approach, but it is possible to obtain a deterministic solution using a quantum algorithm requiring only one oracle query.

	\subsubsection{Sub-problem $ (b) $: Find all secret keys}  \label{sec:subprob_b}
	Next we consider the sub-problem $ (b) $ of the probabilistic Bernstein–Vazirani problem.
	Let $ P(k,m)  $ denote the probability of finding all the secret keys $ s_0 $, $ s_1 $, $ \ldots $, $ s_{k-1} $ in $m$ measurements, or equivalently with $ m $ oracle queries. It is clear that $ 	P(k,m) = 0 $ if $  m < k$. Also, if $ m \geq k $ then it can easily be shown that
	\begin{equation}\label{eq_prob}
		P(k,m) = \frac{1}{k^m} \sum_{\substack{\alpha_1 + \alpha_2 + \ldots \alpha_k = m\\ \alpha_1 \geq 1, \alpha_2 \geq 1, \ldots \alpha_k \geq 1} } \,  \frac{m!}{ \alpha_1! \alpha_2! \ldots \alpha_k!}.
	\end{equation}
	Theorem $ 2.2 $ in \cite{kao1957identity} gives
	\begin{equation}\label{eq_Kao}
		\sum_{\substack{\alpha_1 + \alpha_2 + \ldots \alpha_k = m\\ \alpha_1 \geq 1, \alpha_2 \geq 1, \ldots \alpha_k \geq 1} } \,  \frac{m!}{ \alpha_1! \alpha_2! \ldots \alpha_k!} = 	\displaystyle{\sum_{i=0}^{k-1} \,  (-1)^i \binom{k}{i} (k-i)^m}.
	\end{equation}
	Hence, from \meqref{eq_prob}  and  \meqref{eq_Kao}  we get the following expression for $ P(k,m) $.
	\begin{align}
		P(k,m) = 
		\begin{cases}
			0, \quad &\text{if } m < k, \\
			\displaystyle{\frac{1}{k^m} \,\sum_{i=0}^{k-1} \,  (-1)^i \binom{k}{i} (k-i)^m},  \quad &\text{if } m \geq k.
		\end{cases}
	\end{align}
	For the case of two secret strings, i.e., for $ k=2 $, it follows from the above equation that 
	\begin{equation}\label{eq:prob_two_quantum}
		P(2,m) =  1 - \frac{1}{2^{m-1}}, \quad \text{for }  m \geq 2.
	\end{equation}
	It is clear from the above expression that  $ P(2,m) \to 1 $ (i.e., the probability of finding both the secret keys approaches $ 1 $)  as the number of oracle calls $ m \to \infty $.

	Next consider the classical approach for solution to the sub-problem $ (b) $ of the probabilistic Bernstein–Vazirani problem, i.e., the problem of determining all the secret keys. 
	We note the quantum algorithm for solution to the sub-problem $ (a) $ requires one oracle call to determine any one secret key with probability one. Repeated application of this quantum algorithm will eventually reveal all the secret keys. In other words, the quantum algorithm for solution to the sub-problem $ (b) $ (which involves determination of all secret keys) succeeds with probability $ 1 $ as the number of oracle calls approaches infinity. Indeed, after a large number of oracle calls the sample  probability of occurrence of each secret key from a set of $ k $ distinct secret keys approaches $ \frac{1}{k} $.
	In contrast, even after infinitely many oracle calls the classical approach does not succeed with probability $ 1 $. In fact, with the classical approach,  one can only know the distribution $ r_q $ of the $ q $-th bit, $ 0 \leq q \leq n-1  $, arising from the $ k $ secret keys even after infinitely many oracle calls (on repeatedly querying the oracle with $ x = 2^q $), where $ r_q $ is defined below.
	\begin{equation}\label{eq_def_rq}
		r_q = \sum_{j=0}^{k-1} s_j(q). 
	\end{equation} 
	In other words, $ r_q $ represents the number of secret keys whose $ q $-th bit is $ 1 $.
	For example, consider the case $ n=4 $, $ k= 4 $, with the secret keys with binary representation 
	$ s_0 = 0001  $, $ s_1 = 0011 $, $ s_2 =1011 $ and $ s_3 = 1110 $. In order to get information about the $ 0 $-th bit of the secret keys, one can repeatedly query the oracle with the input $ x = 0001 $ (in binary notation). It is clear that in this case one will observe the output bit to be $ 1 $ with a probability of $ \frac{3}{4} $ as the number of trials (oracle calls) approaches infinity. We note that the output bit contains information about the $ 0 $-th bit of the secret keys. Multiplying the probability ($ \frac{3}{4} $) by the number of secret keys ($ 4 $), one can infer that that $ r_0 = 3 $, i.e., there are three secret keys whose $ 0 $-th bit is $ 1 $. Following a similar approach $ r_q $ can be inferred for the general case. In this particular example, one can also infer that $ r_1 = 3 $, $ r_2 = 1 $ and $ r_3  = 2$. Even after infinitely many trials, the knowledge of $ r_0 $, $ r_1 $, $ r_2 $ and $ r_3 $ will not allow us to determine all the secret keys with high probability. 
	In fact, using the knowledge of $ r_0 $, $ \ldots $, $ r_3 $, and combinatorics, one can determine all the secret key with the probability less than or equal to 
	\begin{equation}\label{eq_special_case}
		\min \left(	\frac{4!}{\prod_{q=0}^{n-1} \, \binom{4}{r_q}}, 1\right).
	\end{equation}
	The above probability bound given in \meqref{eq_special_case} is a special case of a more general result described in \meqref{eq_classical_prob} below.

\begin{figure}[H]
		\centering
		\includegraphics[width=0.9\textwidth]{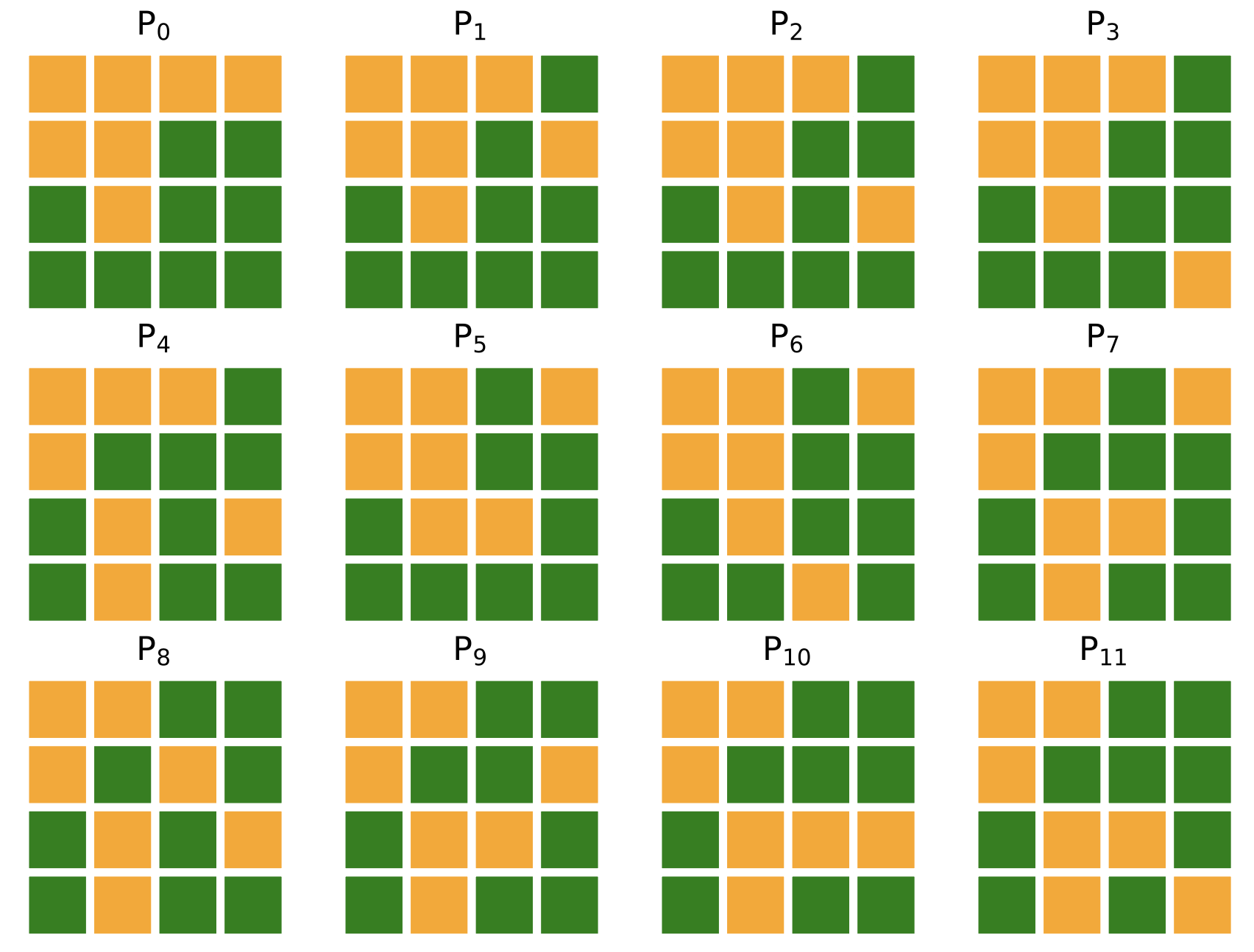}
		\caption{Exhaustive combinations of four keys (each of length four bits) satisfying the constraint 
			($r_3$, $r_2$, $r_1$, $r_0$) = $(2, 1, 3, 3)$, corresponding to the example case discussed in \mref{sec:subprob_b}. 
			With 0 and 1 denoted by orange and green colors respectively, each row of any block ($P_k$) denotes the binary representation of a key.}
		\label{fig:cmatrices}
	\end{figure}

	The exhaustive combinations of keys that satisfy the constraint ($r_3$, $r_2$, $r_1$, $r_0$) = $(2, 1, 3, 3)$ are illustrated in \mfig{fig:cmatrices}. 
	Note each block denotes a combination of secret keys and there are 12 different combinations that satisfy the constraint ($r_3$, $r_2$, $r_1$, $r_0$) = $(2, 1, 3, 3)$. Orange and green colors denote 0 and 1 respectively. Each row of any block denotes the binary representation of a key. While our proposed quantum algorithm for subproblem (b) will succeed in finding the correct combination of all secret keys, shown in top-right block ($P_3$), i.e. $ s_0 = 0001  $, $ s_1 = 0011 $, $ s_2 =1011 $ and $ s_3 = 1110 $, with a high probability (approaching unity, as the number of oracle call/measurements approach infinity), a classical algorithm can be used to only guess the correct combination of secret keys with a low probability of success even after infinitely many oracle calls. 
	For this example, suppose that after infinitely many calls oracle the distribution ($r_3$, $r_2$, $r_1$, $r_0$) is known. Still, on using a classical algorithm, the probability of picking ($P_3$) the correct combination  of all secret keys from a total of $ 12 $ different combinations is $1/12$.  
		
	Assume that after a sufficiently large numbers of oracle calls  the values of $ r_q $ for $ q= 0, \, 1, \, \ldots,\, n-1$ are known. 
	Recall that $ r_q $ represents the number of secret keys whose $ q $-th bit is $ 1 $. It is clear that a total of 
	\[
	\binom{k}{r_q} = \frac{k!}{r_q ! (k-r_q !)}
	\]
	different ways are possible to assign the $ q $-th bit of the $ k $ secret keys such that sum of $ q $-th bit of all the $ k $ secret keys is $ r_q $.  
	Similarly for all the $ n $-bits, (i.e., for $ q=0 $ to $ q=n-1 $), 
	\[
	\mathcal{P} = \prod_{q=0}^{n-1} \, \binom{k}{r_q}
	\] 
	possible choices of picking a set of $ k $ secret keys such that the constraints on the sum of bits is satisfied for all the $ n $ bit positions (i.e., for $ q=0 $ to $ q=n-1 $).
	The observer will not be able to distinguish between these possibilities. In other words, with the knowledge of sum constraints $ r_q $ for $ q= 0, \, 1, \, \ldots,\, n-1$, the observer can infer that there are  $ \mathcal{P} $  possible choices of picking a set of $ k $ secret keys that are consistent with the sum constraints $ r_q $ for $ q= 0, \, 1, \, \ldots,\, n-1$.  Out of all these $ \mathcal{P} $ choices,  at most $ k! $ choices corresponds to the correct secret keys as the order of secret keys does not matter. 
	Therefore, the probability of picking a set of $ k $ secret keys correctly is less than or equal to 
	\begin{equation}\label{eq_classical_prob}
		\min \left(\frac{k!}{\prod_{q=0}^{n-1} \, \binom{k}{r_q}},1 \right).
	\end{equation}
We note that the above upper bound can be made more precise, provided information about identical secret keys is available. 
	If there are $ c $ distinct keys, $ t_0,\, t_1, \,\ldots,\, t_{c-1} $  with  $ \sum_{i=0}^{c-1} b_i = k $, where $ b_i $ denotes the number of occurrences of the secret key $ t_i $, then 
\begin{equation}
R = 	\frac{k!}{ \prod_{i=0}^{c-1} (b_i!)}
\end{equation}
choices corresponds to the correct secret keys, resulting in the following expression for the probability of picking a set of $ k $ secret keys correctly, 
	\begin{equation}\label{eq_classical_prob_precise}
	 \frac{R}{\prod_{q=0}^{n-1} \, \binom{k}{r_q}}.
\end{equation}

	\section{Oracle Implementation} \label{Sec:Oracle}

In this section, we will consider a few examples and different generalizations for oracle implementations.  It is assumed in the following that
the action of the unitary gate $ U_i $, for $ i= 0,\,1, \, \ldots,\, k-1 $,	  is given by
\begin{equation}\label{key}
	U_i \ket{x} \otimes \ket{y} = \ket{x} \otimes \ket{y \oplus f_i(x)},
\end{equation}
where $ f_i(x) = s_i \cdot x$ and the secret key $ s_i \in \{0,1\}^{n}  $.

\subsection{Example: Two secret keys} \label{Sec:two_sec}
\begin{figure}[H]
		\centering
		\includegraphics[scale=1.2]{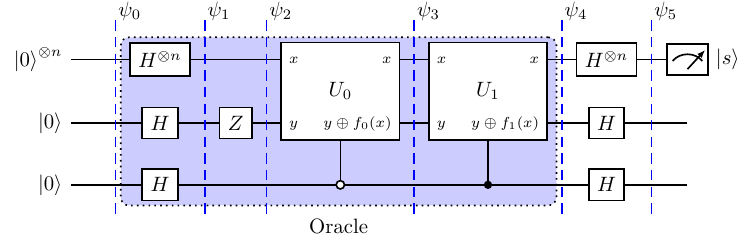}
		\caption{Circuit for implementation of the probabilistic version of the Bernstein-Vazirani algorithm for deciphering two secret keys (each containing $n$ bits)}
		\label{fig:main-2}
	\end{figure}

In order to demonstrate the implementation of the probabilistic oracle, we present an example with two secret keys as shown in \mfig{fig:main-2}. A schematic diagram of the orcale is shown in the shaded color (blue) in  \mfig{fig:main-2}. 
The bottom most ancilla qubit in \mfig{fig:main-2} is used as a control qubit for the unitary gates $ U_0 $ and $ U_1 $. This ancilla qubit is initialized to the state $ \ket{0} $, which then get transformed to the state $ \ket{+} = \frac{1}{\sqrt{2}} \left( \ket{0} + \ket{1}\right)$ upon the action of the Hadamard gate on it. This ensures that the gates $ U_0 $ and $ U_1 $ are selected with equal probability, i.e., $ \frac{1}{2} $ in this case. It can easily be checked that the action of the oracle $ O_2 $ on the top $ n $-qubits (which are initialized to  $ \ket{0}^{\otimes n} $) is given by 
\begin{equation}\label{eq_oracle_two}
	O_{2} \ket{0}^{\otimes n}  =    \frac{1}{\sqrt{2}}  (\ket{\Psi_0}  + \ket{\Psi_1}) 
\end{equation}
where  \begin{equation}\label{eq_phi_i_two}
	\displaystyle {\ket{\Psi_i} = \frac{1}{\sqrt{N}} \, \sum_{x =0}^{N-1} \, (-1)^{f_i(x)} \ket{x}},
\end{equation}
for $ i=0, \,1 $ and $ N = 2^n$. The quantum state given in \meqref{eq_oracle_two} corresponds to the quantum state of the top $ n $-qubits of  $ \ket{\psi_4} $ as shown in \mfig{fig:main-2}. 
shows a complete schematic quantum circuit diagrams for finding two secret keys (each of length $ n $-bits) via the proposed quantum algorithm (described in \mref{sec:pbv_algo}) for solution of the probabilistic version of the Bernstein-Vazirani problem. 
	
\begin{figure}[H]
		\centering
		\includegraphics[width=0.99\textwidth]{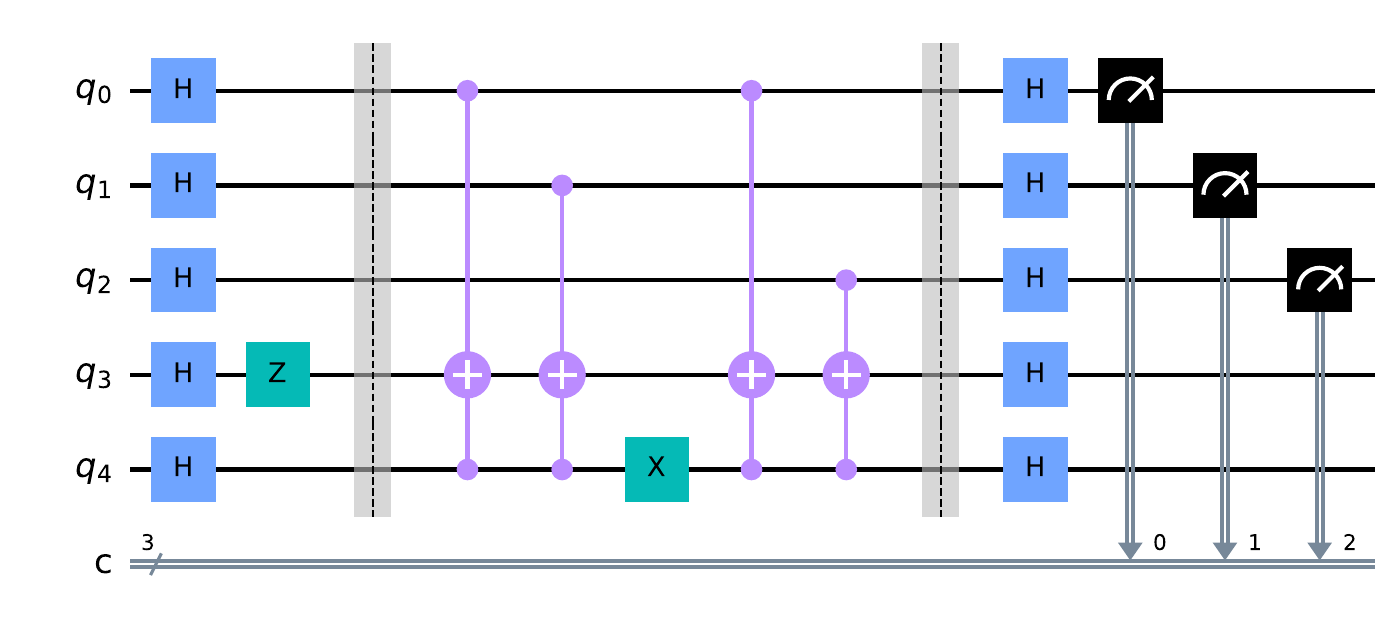}
		\caption{A quantum circuit for the probabilistic version of the Bernstein-Vazirani problem, with details of the oracle implementation, is shown for the case of two distinct secret keys $ (011, \, 101) $. Note that all the input states ($ \ket{q_i} $) are initialized to $ \ket{0} $.}
		\label{fig:circuit_for_two_keys}
	\end{figure}

A quantum circuit for solving the probabilistic version of the Bernstein-Vazirani problem, with details of the oracle implementation, is shown for the case of two distinct secret keys ($ s_0 = 011$ and $ s_1 = 101 $) in \mfig{fig:circuit_for_two_keys}. This quantum circuit was implemented and tested in the simulation environment of IBM's open source platform Qiskit. A histogram representing sample probabilities for selection of the secret keys ($ s_0 = 011$ and $ s_1 = 101 $) is shown in \mfig{fig:histograms_for_two_and_four_keys} on the left (with $ N_{shots}  = 1024$, where $ N_{shots}$ represents the number of repetitions of each circuit for sampling).

	\subsection{Example: Four secret keys }
	
\begin{figure}[H]
		\centering
		\includegraphics[scale=0.97]{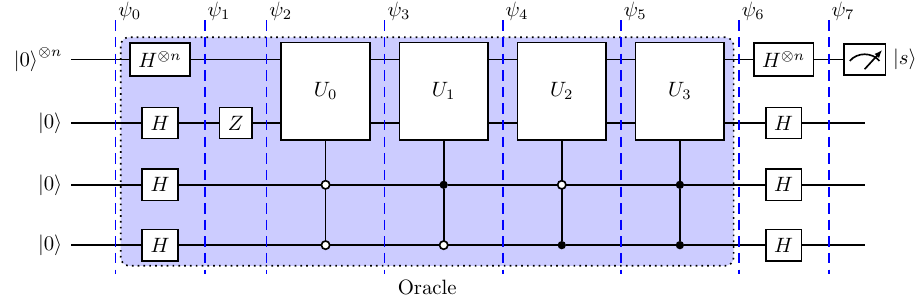}
		\caption{Circuit for implementation of the probabilistic version of the Bernstein-Vazirani algorithm for deciphering four secret keys (each containing $n$ bits)}
		\label{fig:main-4}
	\end{figure}

A similar approach can be used for implementation of the probabilistic oracle  $ O_4 $ with four secret keys as shown in \mfig{fig:main-4}. We note that in this case, we need two control qubits for selecting the controlled unitary gates $ U_0 $, $ U_1 $, $ U_2 $ and $ U_3 $. The bottom two ancilla qubits in \mfig{fig:main-4} are used as the control qubits. One can verify that the action of the oracle $ O_4 $ on the top $ n $-qubits (which are initialized to  $ \ket{0}^{\otimes n} $) is given by 
$ 	O_{4} \ket{0}^{\otimes n}  =    (\ket{\Psi_0}  + \ket{\Psi_1})  + \ket{\Psi_2}) + \ket{\Psi_3})/2 $
where  
$ 	\displaystyle {\ket{\Psi_i} = \frac{1}{\sqrt{N}} \, \sum_{x =0}^{N-1} \, (-1)^{f_i(x)} \ket{x}}, $
for $ i=0,\,1,\,2,\,3 $. This quantum state corresponds to the quantum state of the top $ n $-qubits of  $ \ket{\psi_6} $ as shown in \mfig{fig:main-4}.

\begin{figure}[H]
		\centering
		\includegraphics[width=0.99\textwidth]{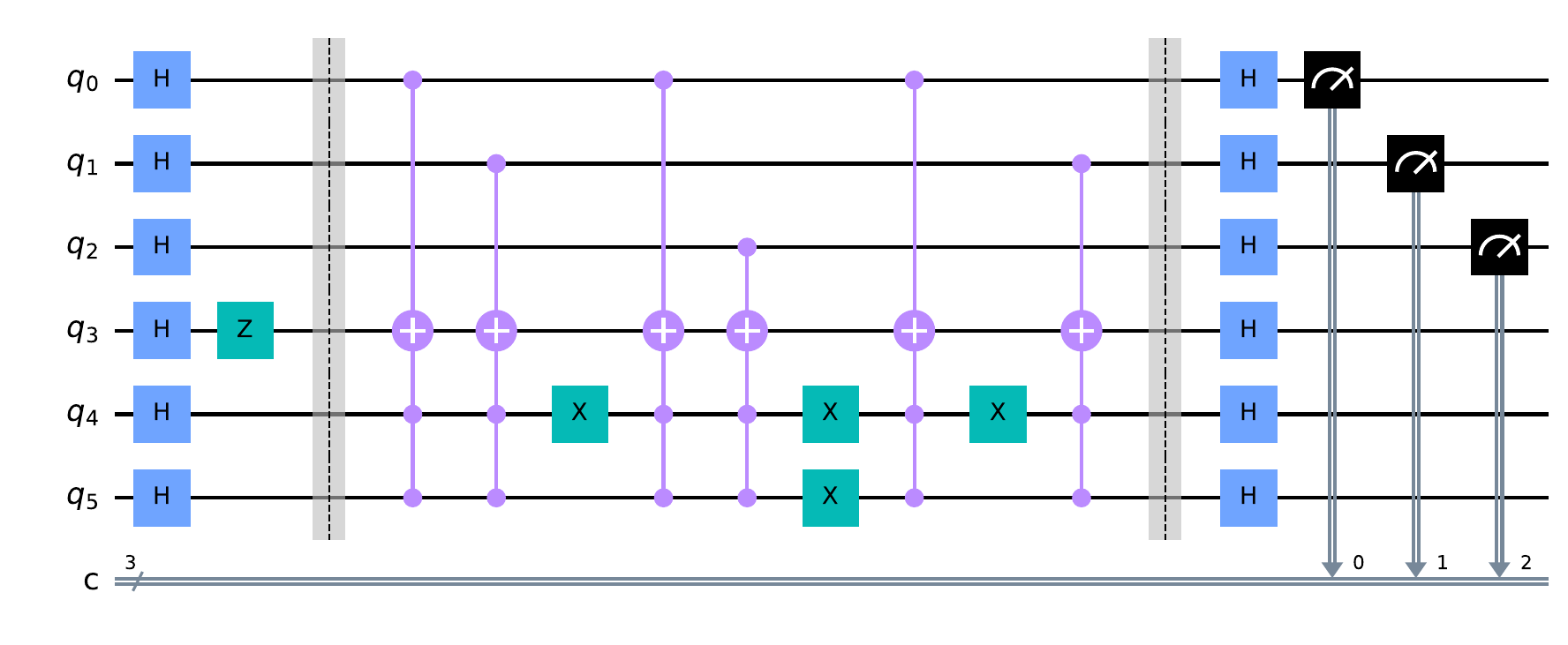}
		\caption{A quantum circuit for the probabilistic version of the Bernstein-Vazirani problem, with details of the oracle implementation, is shown for the case of four distinct secret keys $ (001, 010, 011, 101) $. Note that all the input states ($ \ket{q_i} $) are initialized to $ \ket{0} $.}
		\label{fig:circuit_for_four_keys}
	\end{figure}

\begin{figure}[H]
		\centering
		\includegraphics[width=0.44\textwidth]{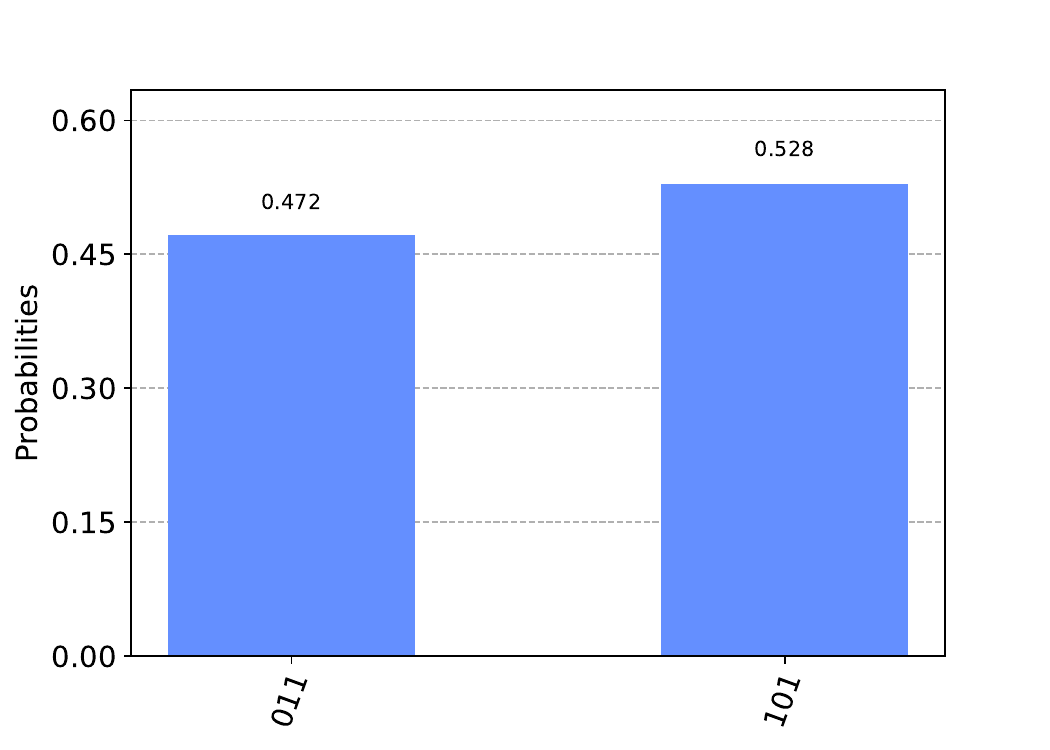}
		\includegraphics[width=0.44\textwidth]{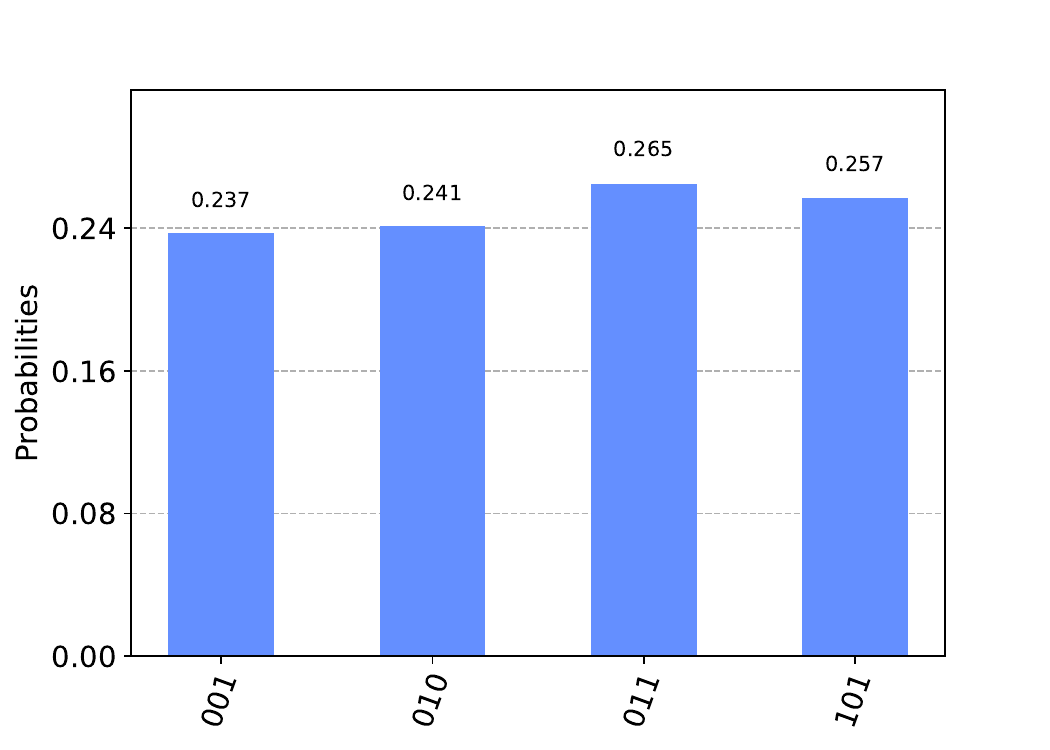}
		\caption{Histograms representing sample probabilities for selection of distinct secret keys. Cases with two secret keys (on left; corresponding to \mfig{fig:circuit_for_two_keys}) and 4 secret keys (on right; corresponding to \mfig{fig:circuit_for_four_keys}) are shown.}
		\label{fig:histograms_for_two_and_four_keys}
	\end{figure}

	A Qiskit-based implementation of a quantum circuit for solution of the probabilistic version of the Bernstein-Vazirani problem  for the case of four distinct secret keys ($ s_0 = 011$, $ s_1 = 101 $, $ s_2 =  011 $ and $ 101 $) is shown in \mfig{fig:circuit_for_four_keys}. A histogram representing sample probabilities for selection of these secret keys  is shown in \mfig{fig:histograms_for_two_and_four_keys} on the right (with $ N_{shots}  = 1024$).

	\subsection{Example: Two identical secret keys out of four secret keys}
\begin{figure}[H]
		\centering
		\includegraphics[width=0.99\textwidth]{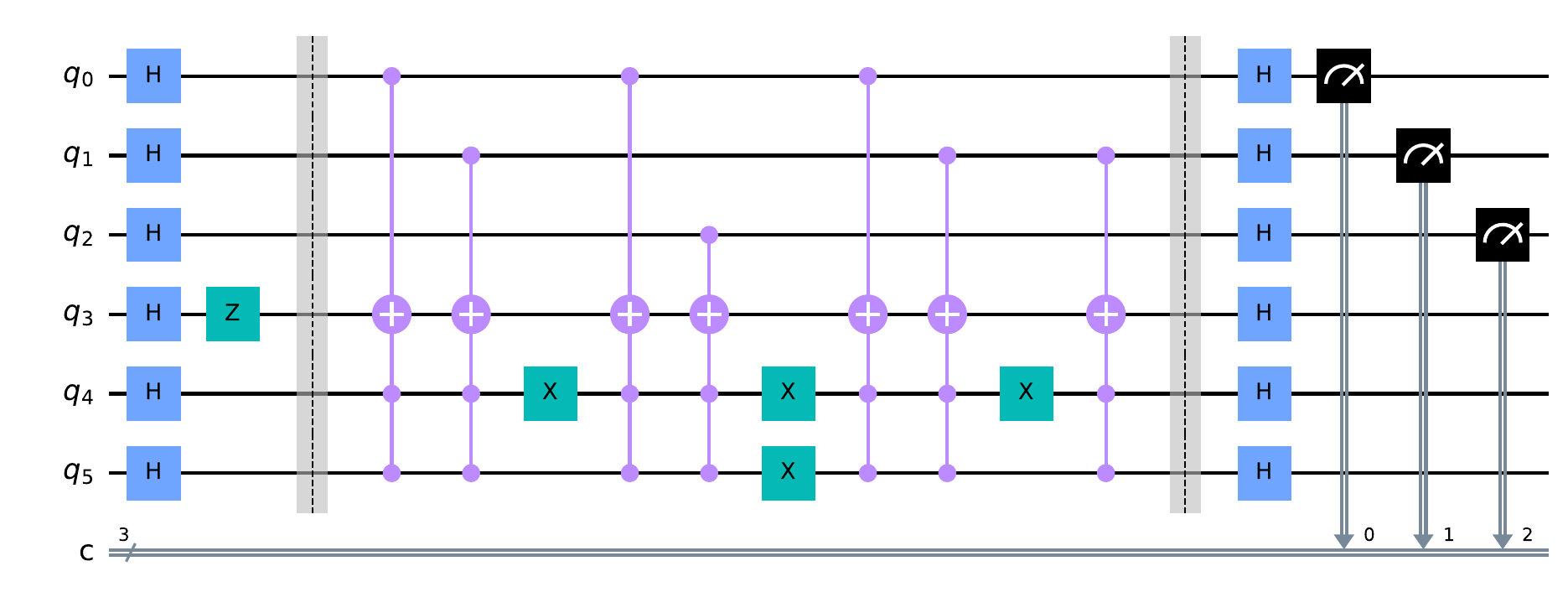}
		\caption{A quantum circuit for the probabilistic version of the Bernstein-Vazirani problem, with details of the oracle implementation, is shown for the degenerate case of four secret keys (010, 011, 011, 101), including two identical keys. Note that all the input states ($ \ket{q_i} $) are initialized to $ \ket{0} $.} 
		\label{fig:circuit_for_four_keys_degenerate}
	\end{figure}

\begin{figure}[H]
		\centering
		\includegraphics[width=0.45\textwidth]{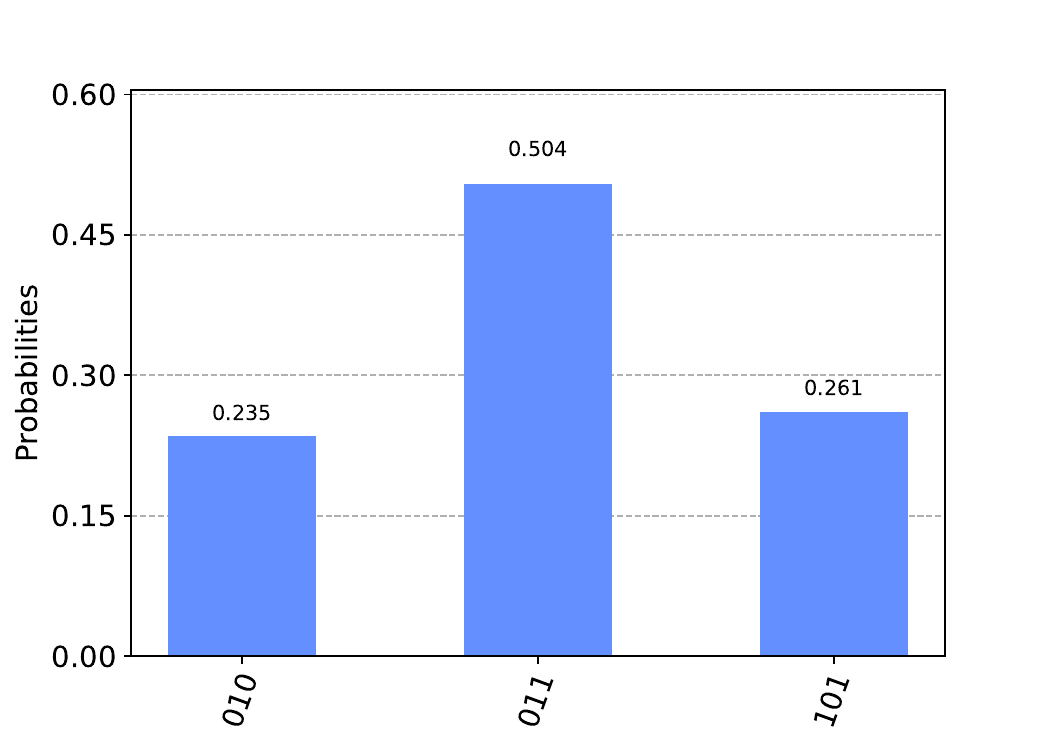}
		\caption{Histogram representing sample probabilities for selection of 4 secret keys (each containing 3 bits), including two identical keys. Corresponding quantum circuit is shown in \mfig{fig:circuit_for_four_keys_degenerate}.}
		\label{fig:histograms_for_four_keys_degenerate}
	\end{figure}

	In the previous two examples cases, the secret keys considered were distinct, which resulted in near uniform sampling of the secret keys. In \mfig{fig:circuit_for_four_keys_degenerate}, we show the Qiskit-based implementation of a quantum circuit for the probabilistic version of the Bernstein-Vazirani problem for the degenerate case of four secret keys $ s_0 = 010 $, $ s_1 = 011$, $ s_2 = 011$, and $ s_3 = 101 $, where two secret keys are identical. The histogram in \mfig{fig:histograms_for_four_keys_degenerate} shows a non-uniform sampling of secret keys, consistent with the expected probability distribution.

	\subsection{General case: $ k = 2^r $ secret keys} \label{Sec:gen_sec}
	The approach used to implement the oracle in the previous examples can be generalized  to implement an oracle  $ O_k $ for the general case of $ k = 2^r $ secret keys (where $ r $ is a non-negative integer), as shown in \mfig{fig:main}. In this case, $ r $ control qubits are needed (the bottom $ r $ ancilla qubits in \mfig{fig:main}), each of which is initialized to $ \ket{0} $. An application of $ H^{\otimes r} $ on these ancilla qubits results in a superposition state $ \frac{1}{\sqrt{k}} \sum_{j=0}^{k-1}  \ket{j} $. This uniform superposition state of $ r $ control ancilla qubits allows the selection of the controlled unitary gates $ U_0 $, $ U_1 $, $ \ldots$, $ U_{k-1} $, with uniform probability. The action of the oracle $ O_k $ on the top $ n $-qubits (which are initialized to  $ \ket{0}^{\otimes n} $) is given 
 in \mref{sec:algo} (e.g., \meqref{eq_oracle} and \meqref{eq_phi_i}). We note that this approach can also be used for the degenerate case in which some secret keys are identical. If any collection of secret keys are identical, then the corresponding unitary gates could be identical. Alternatively, such degenerate cases can be handled more efficiently by appropriate construction of superposition states (or state-preparation) of control ancilla qubits (consistent with the underlying distribution of secret keys). 
	 
	\subsection{General case: $ k \neq 2^r $ secret keys}
	We note that one key step in implementing the oracle (as described earlier) is to create the uniform superposition state $ \frac{1}{\sqrt{k}} \sum_{j=0}^{k-1}  \ket{j} $. A straightforward use of Hadamard gates (in our implementation of oracle) to create the superposition state puts a constraint on $ k $ to be a power of $ 2 $. In case $ k  \neq 2^r$, a version of QFT (Quantum Fourier Transform) algorithm \cite{kitaev1995quantum, mosca2004exact} can be used to create the uniform superposition state $ \frac{1}{\sqrt{k}} \sum_{j=0}^{k-1}  \ket{j} $. 
	Once the control ancilla qubits are in the uniform superposition state $ \frac{1}{\sqrt{k}} \sum_{j=0}^{k-1}  \ket{j} $, they can be used for selection of the controlled unitary gates $ U_0 $, $ U_1 $, $ \ldots$, $ U_{k-1} $, with uniform probability, as described in the previous cases in \mref{Sec:two_sec}--\ref{Sec:gen_sec}. 
	
	\section{Conclusion}

	In this paper, we posed a probabilistic version of the Bernstein-Vazirani problem and proposed a quantum algorithm to solve it. This problem (which represents a generalization of the original Bernstein-Vazirani problem) involves determination of one or more secret keys (encoded in a binary representation) via a probabilistic oracle. 
	
	The proposed quantum algorithm to solve the probabilistic version of the Bernstein-Vazirani problem is found to be superior to any classical algorithm. In order to obtain any single key (with certainty) from a set of multiple keys (subproblem (a)), the quantum algorithm requires only a single query to the (probabilistic) oracle. Further, the proposed quantum algorithm can be used to obtain all keys (subproblem (b)) with high probability (approaching 1 in the limiting case).
	
	In contrast, a classical algorithm will be unable to obtain even a single bit of the secret key with certainty. Even after a sufficiently large number of trials (i.e. even in the limiting case), it is shown that a classical algorithm will be unable to obtain any or all secret keys with certainty. Information relevant to limiting probability distributions of each bit can be obtained using a classical algorithm and this can be used to infer some estimates on the distribution of secret keys based on combinatorics.
	
	It appears that the probabilistic version of the Bernstein-Vazirani problem belongs to an interesting and new class of problems (involving probabilistic oracles), where quantum algorithms can be used to obtain efficient solutions with certainty, where as classical algorithms fail to obtain solutions without inherent uncertainties even in the limiting case (as the number of trials/queries approach infinity).

	\section{Declarations}
	\paragraph{Data availability statement:}
	Data sharing not applicable to this article as no datasets were generated or analysed during the current study.	
	
	\paragraph{Competing interests statement:}	
	The authors have no competing interests to declare that are relevant to the content of this article.

	\bibliographystyle{unsrt}

\end{document}